\documentclass[11pt]{article}

\usepackage[margin=1in]{geometry}
\usepackage{amsmath,amssymb,mathtools}
\usepackage{booktabs,longtable,tabularx,array}
\usepackage{pdflscape}
\usepackage[dvipsnames]{xcolor}
\usepackage{enumitem}
\usepackage{microtype}
\usepackage{makecell}
\usepackage{hyperref}
\usepackage{url}

\hypersetup{
  colorlinks=true,
  linkcolor=MidnightBlue,
  citecolor=MidnightBlue,
  urlcolor=BrickRed
}

\title{From Formulation to Articulation:\\
Considerations for AI-Assisted Research in Information Theory}
\author{Chao Tian and Krishna Narayanan}
\date{}

\begin{document}
\maketitle

\begin{abstract}

Recent advances in artificial intelligence have contributed to several information-theoretic results through genuinely different modes: a language model may suggest a proof, participate in an extended mathematical dialogue, coordinate computation, or operate as one component of a search harness. Treating all of these as “AI-discovered mathematics” obscures distinctions that matter for scientific reliability, attribution, and reproducibility. This article examines a set of recent cases through a four-stage pipeline: how the problem was formulated; how AI participated in finding a solution; how the resulting analytical and computational claims were verified; and how machine-generated outputs were translated into mathematical exposition for human understanding. Based on these cases, we recommend that research publications (1) disclose the nature and extent of AI assistance at each stage, (2) report verification of analytical reasoning separately from certification of numerical computations, and (3) place more emphasis on articulating the underlying ideas, techniques, and interpretations.
\end{abstract}

\section{Introduction}

Information theory is unusually fertile ground for AI-assisted mathematics.
Its problems mix inequalities, probability, discrete optimization, coding
constructions, and computations that can often be certified exactly.
Within this range of tasks, a model might supply a decisive lemma,
generate a proof candidate later reconstructed by researchers, or complete
tedious but otherwise routine derivations. Broad statements such as
``the AI proved the theorem'' obscure these differences.
As AI models become more capable, it is important to examine not only
their contributions to individual results, but also their effects on
the entire research process, from problem formulation to publication.

We view the research process in information theory, from problem formulation to articulation of the results, as a pipeline with four stages. For these stages, we ask the following questions:
\begin{itemize}[topsep=4pt,itemsep=1pt]
  \item \textbf{Problem formulation:} Is the question a longstanding conjecture, a question posed
  first time by the author, or a direction formulated or materially sharpened by AI?
  \item \textbf{AI-mediated solution:} Is the model used through a short prompt, an interactive
  collaboration, or an engineered harness involving search, tools, agents, and repeated evaluation?
  \item \textbf{Verification:} Who checks the analytical reasoning, and what kind of numerical
  certification is supplied?
  \item \textbf{Translation and articulation:} Is the final argument reconstructed and written by human experts, revised from an AI-generated draft, or left close to its machine-generated form?
\end{itemize}
Most of the works discussed below are recent preprints, and both their proofs and their descriptions of AI involvement may change. We assess the research process from public accounts and accompanying artifacts, distinguishing directly supported classifications from inferences. At present, AI is changing the solution stage most directly, with consequential effects on verification and exposition. Whether its influence will extend substantially to problem formulation remains unclear. 

Historically, research publications have served several roles simultaneously: (1) preserving an archival record of scientific progress; (2) presenting sufficient evidence for claims to be evaluated and reproduced; (3) articulating novel ideas, techniques, and interpretations so that researchers can understand and build upon them; and (4) establishing priority and attributing credit. As AI-assisted research becomes more prevalent, these functions may no longer be served entirely by a single conventional paper. AI systems may increasingly evaluate and extend results using detailed records, which include code, computational certificates, interaction histories, and intermediate artifacts. These records may be too extensive or insufficiently organized for direct human consumption. Scientific publications may therefore need to evolve in two complementary directions: preserving a sufficiently complete and machine-readable research record for reproduction and further discovery, while providing a human-oriented exposition that explains the essential ideas, techniques, evidence, significance, and limitations. An important challenge will be to maintain a reliable correspondence between these two layers, so that the human-readable account accurately represents what is supported by the underlying research record. Moreover, as the generation of new results becomes increasingly automated, the traditional emphasis on establishing individual priority and assigning credit for the act of discovery may diminish, while credit for problem formulation, verification, interpretation, and exposition may become more salient.

The human-oriented layer remains important for reasons that extend beyond the immediate production of further scientific results. AI-assisted research is likely to have substantial societal impact in the near future, and society has a legitimate interest in knowing when and how AI systems contributed to consequential scientific developments, how the resulting claims were verified, and to what extent they were understood and endorsed by human experts. The general public relies on researchers and scientific institutions to scrutinize, interpret, and communicate them. The cases studied below reveal a wide spectrum of AI-use patterns and human-review practices, some of which support these purposes more effectively than others. Based on this analysis, we make three recommendations. First, publications should disclose the nature and extent of AI assistance at each stage of the research pipeline. Second, verification of analytical reasoning should be reported separately from certification of numerical computations. Third, reproducibility alone is insufficient: publications must provide a human-oriented articulation of the underlying ideas, techniques, interpretations, significance, and limitations, so that the results can be understood and integrated into broader scientific theory rather than merely reproduced or extended through automated means.

The issues considered here extend beyond information theory and arise
more broadly in mathematical research. Our perspective is influenced
in part by Tao's recent article \cite{tao2026mathematics}, which examines
the implications of AI capabilities for the goals and values of
mathematics under the assumption that systems capable of research-level
mathematical work become available. Our discussion takes a complementary,
case-based approach, examining documented uses of current AI systems
in information-theoretic research. Information theory's longstanding
connections to communication and data compression give these questions
an additional engineering dimension. The practical value of a result
often depends on understanding the coding constructions and design
principles underlying it, beyond the theorem statement itself.
Adapting these ideas to practical constraints requires interpretation
and further engineering investigation, making the articulation of
AI-assisted results particularly important.

\section{From Research Formulation to Articulation: A Taxonomy}

This section introduces a taxonomy for the four stages identified above.  The purpose is to describe
the research process, rather than to assess the importance of a result or assign credit
between humans and AI.  The dimensions are intentionally kept separate: for example, a highly
agentic solution process need not imply weak verification, and a human-written exposition need not
imply that every step was independently checked.  Mixed workflows may exhibit more than one mode;
in such cases, the classification should identify the mechanism that made the material contribution
at each stage.  The taxonomy is defined here without reference to particular works and is applied to
the case studies in Section~4.

\subsection{Problem Formulation}

Problem formulation includes identifying a question, delimiting its scope, and explaining why its
resolution would be consequential.  This stage should be classified according to the origin of the
decisive formulation, not according to the age or difficulty of the problem.  We distinguish three
levels of AI involvement:
\begin{itemize}[topsep=4pt,itemsep=1pt]
  \item \textbf{F-0}: Human-formulated. The target theorem, conjecture, construction, or
  counterexample problem is supplied to the AI system with sufficient specificity that the system's
  task is to seek a solution.
  \item \textbf{F-1}: AI-refined. A human supplies a domain, broad objective, or initial question,
  while the AI system proposes a consequential subproblem, reduction, conjecture, or reformulation
  that materially changes the research direction.
  \item \textbf{F-2}: AI-originated. Given only broad domain-level guidance, the AI system identifies
  a research question and provides a substantive account of its novelty and significance before the
  question is selected for further investigation.
\end{itemize}

The distinction between \textbf{F-1} and \textbf{F-2} depends on both the initial human framing and
the process by which the final question is selected.  Merely restating a human-provided problem,
generating a nearby variant, or supplying motivation after the fact does not constitute an
AI-originated formulation.  An \textbf{F-2} classification therefore requires substantial evidence about the initial framing and subsequent interaction.

\subsection{AI-mediated Solution}

The solution stage concerns how AI contributes after a research question has been selected.  The
phrase ``using an LLM'' spans at least three modes, distinguished primarily by the interaction and
orchestration mechanism:

\begin{itemize}[topsep=4pt,itemsep=1pt]

\item \textbf{S-0}: Direct prompting. A researcher provides the problem, possibly together with
references or a partial derivation, and receives a candidate idea, construction, or proof.  Brief
follow-up prompts may occur, but no sustained collaborative process or external orchestration layer is utilized.

\item \textbf{S-1}: Interactive collaboration. The AI system and human researchers repeatedly
exchange conjectures, examples, objections, computations, and revised arguments.  Humans provide
feedback, select promising directions, detect errors, and reprioritize subtasks as the solution
develops.

\item \textbf{S-2}: Harnessed search or agentic workflow. The AI system operates within a structured (automatic iteration)
loop involving one or more components such as program execution, evaluators, search or population
management, external tools, multiple agents, proof-state tracking, or automated iteration.

\end{itemize}

These modes also reflect the nature of the researcher's involvement. A direct prompt may follow substantial human work in formulating a precise question, while an interactive collaboration may require researchers to supply intermediate insights, diagnose failed arguments, and guide successive attempts. A harness shifts much of the researcher's effort toward designing the search space, evaluators, and computational workflow. The difficulty and structure of the problem therefore influence which mode is useful and how much human reasoning it requires. In assessing a result, the solution mode should be considered alongside the documented human contributions and the balance between conceptual reasoning and computation; none of these alone determines intellectual credit or reliability. Reporting such mode will help fellow researchers who would follow up on the reported research results, either to reproduce, or to develop further results. 

\subsection{Verification}

Verification concerns whether the published claim is adequately supported, independently of how the
candidate solution was generated.  AI assistance makes the verification process especially important
to document, but it does not by itself transfer responsibility away from the authors of the published
work.  A further distinction is essential: checking the analytical reasoning is different from
certifying a finite computation (such as a finite precision certificate) produced after an analytical reduction. 
We use the following scale for the analytical-reasoning component:

\begin{itemize}
  \item \textbf{A-0}: No analytical verification. The indispensable analytical reasoning has not been completely
  checked by human experts or a proof assistant.   
  \item \textbf{A-1}: Author or collaborator verification. The authors or mathematically qualified collaborators involved in developing the argument checked or reconstructed the complete reasoning chain. Merely accepting responsibility for the paper does not by itself establish this level.
  \item \textbf{A-2}: Independent expert verification. A qualified expert outside the proof-generation
  loop checked the submitted reasoning and its indispensable reductions.
  \item \textbf{A-3}: Formal verification. The analytical argument has been formalized in Lean or another
  sound proof assistant and checked relative to a trusted base.
\end{itemize}

In traditional research papers, author verification is generally presumed: the
authors developed the result, present its proof, and take responsibility
for its correctness. This convention corresponds broadly to \textbf{A-1},
although it does not tell readers precisely which steps the authors checked
or how they checked them. When an independent expert has examined the indispensable steps of a particular proof, that contribution may support
\textbf{A-2}; an acknowledgment should make the scope of the examination clear. 
With AI-assisted research, these two levels are retained, and should be made more explicit in the publication. 

The distinction between \textbf{A-0} and \textbf{A-1} becomes especially
important when AI systems generate substantial parts of an argument. 
Except in exceptional circumstances, publication of a claimed
theorem should require a complete check of its indispensable
analytical reasoning by the authors or qualified human experts,
or formal verification of the corresponding claim
(\textbf{A-1}, \textbf{A-2}, or \textbf{A-3}).
Authors should state which checks were performed. 
A rigorous numerical
certificate cannot replace this requirement when the reduction
connecting the certificate to the theorem has not been verified.

Formal verification
(\textbf{A-3}) offers another route to establishing correctness, including
when the authors cannot independently check every step of a long proof.
It establishes the formalized claim relative to its stated definitions,
assumptions, and trusted base. The correspondence between that claim and
the theorem described in the paper must still be checked, which can be a non-trivial task, and the library 
dependence in such formalized proofs (e.g., in Lean) may introduce hidden issues difficult to debug. Moreover, a
formally verified claim does not by itself explain the ideas behind the
result or support additional assertions made only in the human-readable
exposition.

For results that depend on numerical optimization, root solving, exhaustive search, or a computationally verified counterexample, we use a separate numerical scale:
\begin{itemize}
  \item \textbf{N-0}: Not used or not applicable. The proof is completely analytical, or numerical work is only
  exploratory and carries no logical load for the correctness of the result.
  \item \textbf{N-1}: Ordinary numerical evidence. Floating-point experiments, Monte Carlo results,
  or plots provide supporting evidence but do not constitute a rigorous certificate
for the relevant claim. 
  \item \textbf{N-2}: Rigorous certificate. A precisely stated finite claim is established using exact rational
arithmetic, interval or ball arithmetic with outward rounding, an exhaustive finite check, or another rigorously checkable
computational certificate. 
  \item \textbf{N-3}: Independently replayed or formally integrated certificate. The certificate was
  reproduced independently, or its checker and connection to the theorem were formally verified.
\end{itemize}

For example, a computer-assisted proof may follow the workflow
\begin{equation}
  \text{theorem claim}
  \xrightarrow{\text{analytical reduction}}
  \text{finite obligations}
  \xrightarrow{\text{certified computation}}
  \text{verified obligations}.
\end{equation}
An \textbf{N-2} certificate establishes only the second step.  If the analytical reduction remains at
\textbf{A-0}, the overall theorem remains provisional.  More generally, the overall assessment is
controlled by the weakest indispensable link, rather than by the strongest artifact in the work.

\subsection{Translation and Articulation}

The final stage concerns how machine-generated ideas, derivations, computations, and artifacts are
converted into an account intended for human understanding.  Translation and articulation mean more
than stylistic editing: reconstruction can expose missing quantifiers, hidden regularity assumptions,
or a mismatch between a program and the theorem it is supposed to establish.  We distinguish three
modes according to the degree of human reconstruction:

\begin{itemize}
  \item \textbf{T-0: Machine-led presentation.} AI produces the final exposition, with human involvement limited to direction, feedback, or a brief final review.
  \item \textbf{T-1: Author-revised AI exposition.} Substantial prose or derivations originate in AI
  output, while the authors review, edit, reorganize, and clarify the material for publication.
  \item \textbf{T-2: Human-led articulation.} Human researchers construct or substantially reorganize the final
exposition around the underlying ideas, techniques, interpretations, significance, and limitations. AI may have contributed to the
mathematics or assisted with drafting. This classification concerns the presentation and does not, by itself, establish analytical
verification.
\end{itemize}

A formal proof or machine-readable research record is complementary to this scale and should be
described through the verification and reproducibility record; it does not by itself provide
human-oriented articulation.  Conversely, polished exposition does not establish correctness.
Translation may contribute to author verification by exposing errors during reconstruction, but the
\textbf{T}- and \textbf{A}-levels should remain separately reported.

It should be noted that even in traditional research and publications, the same result may also be presented and published in multiple forms, e.g., first as a research technical article, and then later as an expository or tutorial article. With AI-assisted research, the
distinction between recording a result and explaining its underlying
ideas may become more pronounced, because a detailed machine-generated
derivation may be available before a satisfactory account for human
understanding can be developed.

\section{Granularity and Attribution of AI Contributions}

Two related questions arise when the contributions of humans and AI
systems are intertwined: what should be treated as the unit of analysis,
and whether those contributions should be quantified.

\subsection{Hybrid Work with Distinct Components}

The classifications in Section~2 need not apply uniformly to every
part of a research article. At one extreme, human researchers may
formulate a theorem and develop most of its proof, while using AI to
establish one lemma or overcome a decisive obstacle. At the other
extreme, a largely AI-generated argument may depend on a crucial
human reduction or correction. Describing either work simply as
``human-proved'' or ``AI-proved'' would obscure its development and
make it harder for future researchers to identify and build on the
principal ideas.

When distinct contributions can be identified, the appropriate unit
of analysis may be a substantive component, such as a reduction,
lemma, construction, counterexample, proof, or finite certificate.
Authors should describe the principal components developed with AI
assistance, the preceding mathematical work relevant to understanding
that assistance, and how those components were incorporated into the final result. A short
lemma may warrant prominent disclosure if it resolves the principal
obstacle to a theorem; its length in the finished paper does not measure
its importance.

In a sustained interactive collaboration, the development of individual
components may be too intertwined to reconstruct in this way. A
paper-level account may then be more appropriate, provided it describes
the respective roles of the researchers and the AI system and identifies
any decisive contributions that can be distinguished. The aim is an
informative account of how the result was obtained, not an exhaustive
attribution of every proof step.

Origin and verification must be kept distinct. If AI supplied a key
lemma and the authors subsequently checked both its proof and its role
in the theorem, the lemma remains an AI-assisted contribution, while
analytical verification of the complete argument may reach
\textbf{A-1}. Human verification does not change the origin of the
argument. Conversely, a human-produced proof that relies on an
unchecked AI-generated lemma does not reach \textbf{A-1} for the
complete theorem.

\subsection{Quantification of Contributions}

Should a percentage split between human and machine contributions be indicated? We believe this can be extremely difficult to quantify, and also unnecessary. Such a number lacks a stable interpretation
unless the quantity being measured is specified. Time spent, amount
of text, number of proof steps, originality, and importance to the
result can yield very different estimates, some of which can be highly subjective measures. 
Contributions may also be interdependent: a human-formulated reduction may enable an AI-generated
lemma, which in turn enables the remainder of a human proof. 

We aim to characterize the research process, rather than
to allocate intellectual credit or reconstruct the full history of
each result. Earlier mathematical work and human preparation are
relevant to this purpose when they clarify the task presented to AI, the information
and guidance supplied, or the steps needed to integrate its output
into the final argument. A detailed account of who originated every
preceding idea is generally unnecessary for this purpose. Such
questions remain important to scholarly attribution, but they are
not specific to AI-assisted research and lie outside the scope of
our classification framework. 



We therefore favor disclosures that explain the respective roles
of humans and AI systems and identify substantive components
developed with AI assistance, where these can be distinguished.
The aim is to make the workflow intelligible without requiring
a division of credit among all contributors to its mathematical
foundations. Authors may additionally explain which contributions
they consider decisive, including earlier work that enabled AI-assisted discovery. 
Such information may be valuable for future research on related problems. 
In this note, however, we discuss such contributions only where they clarify the workflow; 
assessing their relative intellectual importance lies outside our scope. 
The reasoning/computation profile in the next section describe the
character of the mathematical work; they do not assign shares
of that work to humans and AI.

\subsection{Documented vs. Inferred}
\label{subsec:documentationandinference}

For this note, we mostly assign the classifications based on publicly available disclosures or explicitly stated approaches 
provided in the papers and the accompanying artifacts (certificates, program files, data, or interaction records). 
Some assignments, however, require an inference about aspects of the workflow that are not explicitly
described. When it is reasonably obvious from context (based on our expert judgment) but not clearly stated in the authors' disclosure, we mark with an asterisk, such as
\textbf{T-2*} to indicate it is an inferred classification. Such assessments necessarily involve some subjectivity. This distinction is particularly relevant to translation and
articulation, since authors often describe AI-assisted discovery
and verification with more details than the preparation of the final
exposition. This partially reflects the fast-changing landscape of AI-assisted research, where translation and articulation
are yet to be viewed as a significant or necessary component of disclosure.

\section{Case Studies}

We now apply the taxonomy to public descriptions of recent AI-assisted results identified through an AI-assisted literature search of arXiv preprints listed in cs.IT and posted before October 25, 2026; the selected cases are illustrative rather than exhaustive. For each work, we assess the latest publicly available version before the same date. We also include two manuscripts from OpenAI’s public mathematics collection \cite{openai_math_2026}, which are not hosted on arXiv. We organize the cases primarily by the documented mode of AI participation,
distinguishing work with limited workflow disclosure, interactive mathematical
development, and harnessed or agentic approaches. A final subsection compares
the different workflows used on the Courtade--Kumar conjecture and related
problems.  Inferred classifications are marked with an asterisk, following
Section \ref{subsec:documentationandinference}.

\subsection{AI-Assisted Discoveries with Limited Workflow Disclosure}

\paragraph{Gaussian completely monotone conjecture.}
Gu and Sellke \cite{gu_sellke_2026} give an explicit probability
measure whose Gaussian-smoothed entropy violates the relevant
complete-monotonicity property, disproving the Gaussian completely
monotone conjecture posed about ten years earlier
\cite{cheng2015higher}. It was reported that GPT-5.5 Pro found the
counterexample. The published object is then analyzed by conventional
argument, with SageMath/Arb computations using exact inputs and
rigorous ball arithmetic. The public disclosure does not establish
whether the model produced the example after a direct prompt or a
longer interaction. We infer a classification of \textbf{T-2*} as an assessment of human-led articulation rather than an explicit writing disclosure.

\paragraph{Fisher information along the heat flow.}
Zou et al.\ construct a smooth positive density in two dimensions
that violates log-convexity of Fisher information along heat flow
\cite{zou_et_al_2026}, disproving the same complete-monotonicity
property in higher dimensions. Their revised abstract states that the counterexample was found by 
GPT-5.5 Pro. The final proof uses a hexagonal
perturbation on a triangular torus and an analytical transfer to a
Gaussian-decaying density; the Fourier-grid calculation provides
supporting evidence rather than carrying any obligation of the proof. 
The precise interaction with the model is
not disclosed. We infer a classification of \textbf{T-2*} as an assessment of human-led articulation rather than an explicit writing disclosure.

\paragraph{The undirected multiple-unicast conjecture: Reed-Solomon local codes.}
Braverman and He~\cite{braverman_he_2026} construct an undirected network
in which linear network coding outperforms fractional multicommodity
flow, therefore disproving the conjecture posed in \cite{li_li_2004}. 
GPT-6 assisted in finding a nondeterministic counterexample using
Reed--Solomon local codes; an edge orientation and local search then
converted it into a causal communication scheme. The disclosure credits
ChatGPT with the counterexample and proof drafts and explicitly states
that the authors checked the proof, supporting \textbf{A-1}. The account
combines mathematical construction with computational exploration. The accompanying exact symbolic verifier checks decoding for all source inputs by linearity, causal availability, and terminal distances,
supporting \textbf{N-2}. The disclosure does not sufficiently specify the interaction and
expository workflows to assign \textbf{S}- and \textbf{T}-levels.

\paragraph{The undirected multiple-unicast conjecture: reversible-circuit approach.} A concurrent work \cite{zhang_li_li_2026} gives a different counterexample to the
undirected multiple-unicast conjecture, using a reversible-circuit
construction, and extends it to a family with a growing coding
advantage. The authors disclose that GPT-6 Astra assisted in developing
the proofs and Lean code, but do not describe the interaction or
expository workflow in sufficient detail to assign \textbf{S}- or
\textbf{T}-levels. The published proof is analytical, with no indispensable
computational certificate, supporting \textbf{N-0}. They report Lean formalization of both the finite
counterexample and the family theorem, supporting \textbf{A-3}. 

\paragraph{Three conjectures for the doubly symmetric binary source.}
Pichler~\cite{pichler_2026} proves, for binary auxiliary variables, an averaged binary-symmetric-channel conjecture and, at $p=0$, two double-sided information-bottleneck conjectures. The proofs were developed with
substantial LLM assistance; Claude Opus 5 produced most of the Lean
development, and Claude Fable 5.1 contributed a polynomial-certificate
check. All three main theorems are reported as checked by Lean using
standard axioms (\textbf{A-3}). The maximization result contains formally
integrated polynomial and interval certificates (\textbf{N-3}); other
results without numerical proof obligations are \textbf{N-0}. 
The formalization preceded the substantially AI-assisted drafting of
the paper for human readers. This sequence is relevant to articulation,
but the disclosure does not establish a specific \textbf{S}-level or \textbf{T}-level.

\paragraph{A formally verified deletion-channel upper bound.}
Soysal~\cite{soysal_2026} proves that binary deletion-channel capacity
satisfies $C(d)\leq(1-d)/4$ for $13/20\leq d<1$. The argument combines
backward output descriptions, an entropy identity, a finite-window
potential, and deletion composition. Claude Fable 5.1, GPT-5.6 Sol,
and GPT-6 Astra are named as development tools, but the disclosure
does not specify their separate tasks or interaction protocol.
The paper reports an exact integer certificate and end-to-end Lean
formalization, supporting \textbf{A-3}/\textbf{N-3}. This is a separate
result from Papailiopoulos's capacity approximation. The brief process
disclosure does not support specific \textbf{S}- or \textbf{T}-levels.

\paragraph{Recursive constructions for zero-error capacity.}
Tandon~\cite{tandon_2026} improves an odd-cycle Shannon-capacity lower
bound through heterogeneous recursive constructions. Intermediate
objects with the same current code size can differ in their usefulness
in later recursion, and the method exploits those differences.
ChatGPT and Claude assisted with mathematical ideas, proofs, searches,
verification scripts, and exposition. Exact arithmetic and finite
checks support the construction bounds (\textbf{N-2}).
The disclosure accepts author responsibility without explicitly
attesting to a complete human analytical verification, therefore, does not offer sufficient details to determine the \textbf{A}-level. 
The interaction mechanism and extent of human
revision are insufficiently specified to assign \textbf{S}- and \textbf{T}-levels.

\paragraph{Explicit constructions for odd-cycle Shannon capacity.}
Itty et al.~\cite{itty_et_al_2026} improve lower bounds for Shannon
capacities of several odd cycles through independent sets in strong
graph products. The constructions were obtained through multiple
prompts to ChatGPT-5.6 Sol Pro; one component also involved local search
executed by the model. It is not clear what the dominant interaction is. The authors explicitly report verifying the constructions'
feasibility and writing the manuscript themselves, with AI used only
for minor language edits. The reported feasibility checks support \textbf{N-2}, and the
author-written manuscript supports \textbf{T-2}. These statements
do not by themselves establish the scope of analytical verification. 

\subsection{Interactive and Guided Proof Development}

\paragraph{Bernoulli measures conditioned on affine subspaces.}
Wu resolves an Ancheta--Massey-type entropy question for Bernoulli
parameters below one half \cite{wu_bernoulli_2026}. The paper states
that the proof was discovered by GPT-5.6 Sol under interactive
guidance. It also records the translation step: the author simplified
the program, developed a self-contained proof, wrote the paper, and
accepted responsibility. This is an example of an AI-assisted route
that the human author subsequently reconstructed. The proof is
essentially analytical; we assess verification as \textbf{A-1}/\textbf{N-0}. A \textbf{T-2} classification is supported by the described reconstruction and writing effort.

\paragraph{Marton's inner bound.}
Huang et al. showed that Marton's inner bound is not
generally optimal for the two-receiver broadcast channel
\cite{marton1979coding,huang_liu_liu_2026}. The acknowledgments
describe Chandra Nair proposing an AI-assisted counterexample search
at ISIT 2026. Several model-assisted stages followed, including a
Markovity counterexample, geometric elimination, ternary examples,
and refinements from constrained to unconstrained statements. The
reported models include GPT-5.6 Sol and Claude systems. Nair and 
Gohari are reported to have checked technical nonnumerical
aspects and also provided guidance during the investigation. The reported expert checks support analytical verification, while
rational distributions and interval arithmetic support
\textbf{N-2}. Classification as \textbf{A-2} additionally requires
that the relevant check be independent of the proof-generation
process, yet this was not made clear in the disclosure, and we therefore record it as \textbf{A-1/A-2}, indicating the uncertainty. A \textbf{T-2*} classification is inferred, as 
assistance with organization and writing is documented, but the full division of expository work is less explicit. 

\paragraph{Second-order asymptotics for the Gaussian MAC.}
In two connected preprints, Tan characterizes the second-order
coding-rate region of the two-user Gaussian multiple-access channel
at its corner points \cite{tan_mac_corners_2026} and at the relative
interior of the sum-rate face \cite{tan_mac_face_2026}. The latter
work shows that, for error probabilities below one half, a suitable
mixture of power splits improves on constant-power inner bounds, and
establishes a matching converse.

Both manuscripts disclose substantive assistance in exploring and
stress-testing proof strategies, formulating and refining intermediate
lemmas, checking calculations and possible gaps, and refining
exposition. The corner-point paper names OpenAI Codex (GPT-5.6 Sol);
the subsequent paper names OpenAI Codex (GPT-6 Astra) and Claude
(Opus 5). Tan states in each manuscript that he independently
verified every mathematical statement and proof in the final version.
This is unusually explicit documentation of \textbf{A-1}
verification. The disclosures do not isolate a particular
AI-generated lemma or give enough interaction history to quantify
the model's contribution to each proof. The results are predominantly
analytical; their illustrative numerical work does not appear to
carry an indispensable proof obligation (\textbf{N-0}). We infer \textbf{T-2*}
from the public account and the 
presentation of the result, although the division of work in
preparing the final exposition is not explicitly described.

\paragraph{Exact-repair regenerating codes.}
Tian~\cite{tian_regenerating_2026} characterizes the storage--repair-bandwidth
tradeoffs for $(6,5,5)$ and $(7,6,6)$ exact-repair regenerating codes.
In an author-guided investigation, GPT-6 Astra proposed a conditioning
argument based on shortened codes and carried out restricted searches
over entropy-linear-program constraints. Further prompting produced a
submodularity-based argument, exact integer certificates, a Python
verifier, and an initial manuscript. The author reports verifying the
reasoning, revising the exposition, and reviewing and testing the
verifier, supporting \textbf{A-1}/\textbf{N-2}. We classify the interaction
as \textbf{S-1} and the documented revision of an AI-generated draft as
\textbf{T-1}. The case combines a structural reduction with substantial
certificate search.

\paragraph{Sharp Rogers--Shephard inequalities for entropy.}
Han et al.~\cite{han_et_al_2026} establish sharp weighted
Rogers--Shephard inequalities for R\'enyi entropy of differences of
independent copies of the same log-concave random vector, including the Shannon endpoint. They report extensive assistance from ChatGPT, Codex, and Kimi in exploring proof strategies and counterexamples, developing derivations, 
and drafting and revising the manuscript. The authors explicitly state
that they reviewed and verified all final mathematical arguments,
supporting \textbf{A-1}. We assess this as interactive proof development
(\textbf{S-1*}) with author-revised AI exposition (\textbf{T-1*}). The
proofs are analytical (\textbf{N-0}).

\subsection{Harnessed Search and Agentic Proof Production}

\paragraph{Deletion-correcting codes.}
Weindel and Heckel adapt a FunSearch-style evolutionary loop to
deletion-correcting codes \cite{weindel_heckel_2025}. An LLM
repeatedly proposes scoring or construction programs; a
conventional evaluator ranks them and retains useful variants.
For one deletion, a discovered function is converted into a
theorem recovering a conjectured-optimal Varshamov--Tenengolts code.
For more deletions and larger alphabets, the gains are principally
improved constructions and empirical heuristics. The paper also
reports that sampling more candidate functions was more useful
than allocating the same resources to longer reasoning traces.
The workflow is predominantly search and optimization with an analytical theorem extracted
for one part of its output. A \textbf{T-2*} is inferred based on the assessment 
that humans extracted and articulated the mathematical result.

\paragraph{Binary deletion-channel capacity.}
Papailiopoulos presented explicit upper and lower bounds for 
binary deletion-channel capacity, leading to an approximation of the capacity to within one hundredth of a bit
\cite{papailiopoulos_2026}. The project used GPT-5.6 Sol,
GPT-6 Astra, and Claude Fable in coding-agent environments.
According to the disclosure, agents developed arguments, designed
and ran numerical checks, reviewed proofs, drafted and revised the
manuscript, and exchanged intermediate results; the author posed
the question, steered the direction, and directed the presentation.

A repository provides large proof inputs and code for exact or
interval checks \cite{papailiopoulos_repo_2026}, which supports \textbf{N-2}. In his
account of the project, the author describes agents auditing
one another and states that he did not fully understand the
proof~\cite{papailiopoulos_blog_2026}. On that basis, we infer
\textbf{A-0}$^{*}$ for the proof as released. We view 
the overall claim as a provisional
AI-generated computer-assisted proof with certified numerical
premises: the certificates cannot by themselves verify the
reduction connecting those premises to channel capacity. A \textbf{T-0} classification is appropriate, 
supported by the described agent-generated manuscript and author direction. 

\paragraph{Fourier entropy of threshold functions.}
Shmalo~\cite{shmalo_2026} presents a proof of the Fourier entropy--influence
inequality for all linear threshold functions, together with extensions
to compositions and product measures. The disclosure describes proof
development with Claude and GPT-6 Astra Pro using collaborating agents,
followed by model-based reviews of arguments, constants, and finite
computations. AI also assisted manuscript preparation. This is an
agentic workflow (\textbf{S-2}). The author reports directing the work and accepting
responsibility, but does not document a complete human
analytical check or formal verification; we therefore leave the \textbf{A}-level unassigned. Model audits do not establish independent expert
verification. The rigor and indispensable scope of the finite
computations, and the extent of human reworking of the exposition,
require clarification before assigning N and T.

\subsection{Comparing workflows on the Courtade–Kumar and related problems}
\paragraph{Courtade--Kumar conjecture: the entropy-production and spectral approach.}
Ky and Tran~\cite{ky_tran_2026} give a proof of the Courtade--Kumar conjecture by combining entropy
production, Fourier estimates, and spectral arguments. Their disclosure
states that they formulated the approach and used ChatGPT to develop
proofs, prepare verification code, and revise the exposition. The
interaction protocol is not
specified; therefore, we leave the \textbf{S}-level unassigned. Appendix A of \cite{ky_tran_2026} documents 
the exact rational checks, outward-rounded Arb computations, and verification of
domain coverage, supporting an \textbf{N-2} classification. The disclosure accepts
authors' responsibility but does not explicitly attest to a complete
human analytical check, therefore, we leave \textbf{A}-level unassigned. We infer \textbf{T-2*}
from the public account and the 
presentation of the result.

\paragraph{Courtade--Kumar conjecture: the differential-equation and Bellman approach.}
Chen et al.~\cite{chen_et_al_ck_2026} give another independent proof of the
Courtade--Kumar conjecture using a differential-equation reduction and
Bellman inequalities. Their disclosure describes sustained mathematical
collaboration with Gemini in the Stellar Colosseum multi-agent harness,
with additional contributions involving Anthropic and OpenAI models.
AI systems developed analytical arguments, partitioned parameter regions,
and carried out interval-arithmetic computations under human guidance.
The workflow combines \textbf{S-1} and \textbf{S-2}. The September 24
revision reports end-to-end Lean verification of the proof, including
all numerical certificates, supporting \textbf{A-3}/\textbf{N-3}.
A separate expository note explains the reduction and key lower bounds. 
We infer \textbf{T-2*} for the accompanying expository
account.

\paragraph{Courtade–Kumar conjecture: entropy flow and multibit extensions.}
Mahdavifar and Beirami developed another independent proof of the Courtade--Kumar
conjecture for every dimension and binary symmetric-channel
parameter, together with multibit extensions
\cite{mahdavifar_beirami_2026}. Their disclosure distinguishes
two components of the same paper that should not receive a single
description of AI involvement.

For the \emph{single-bit theorem}, the disclosure states that generative AI
systems produced the proof and its detailed derivations under the
authors' high-level direction and feedback. The argument combines entropy flow, boundary inequalities,
Fourier analysis, and logarithmic Sobolev estimates. In the revised manuscript \cite{mahdavifar_beirami_2026}, the
authors link a Lean~4 formalization of their one-bit proof,
supporting \textbf{A-3}. The accompanying numerical calculations are described as supplementary
diagnostics rather than proof inputs, supporting \textbf{N-0}. A \textbf{T-0*} is suitable given the reported generation-and-review workflow.

For the \emph{multibit extension}, the authors directed the multibit investigation, 
selecting constructions and computational questions, and interpreted the evidence and
conjectures. They construct a shortened-Hamming-code quantizer that beats the
coordinate benchmark for $k=8$ output bits; a direct-sum argument
extends this failure of coordinate optimality to every $k\geq8$.
For $2\leq k\leq7$, searches within selected structured-code families
and Monte Carlo experiments found no counterexample, motivating
a conjecture rather than establishing optimality. This component is thus a
human-directed hybrid of construction, proof assistance, and
computational exploration. Its proved construction claims and its
numerically motivated conjecture should be assessed separately;
evidence from Monte Carlo search does not certify the conjecture. We infer \textbf{T-2*}
from the public account and the 
presentation of the result, although the division of work 
in the exposition is not explicitly described.
The counterexample for each $k\geq 8$, valid for crossover
probabilities sufficiently close to $1/2$ from below, follows
analytically (\textbf{N-0}). Below eight bits, the reported
exact enumeration establishes a maximum within a specified
seven-bit family (\textbf{N-2}); the broader numerical
searches provide evidence for the small-output conjecture
(\textbf{N-1}).

\paragraph{A weak Hellinger inequality for noisy Boolean channels.}
Durcik, Fraccaroli, and Roos~\cite{durcik_et_al_2026} prove a weak
Hellinger inequality by reducing it to an inequality in three real
parameters. GPT models contributed substantially to the latter proof
through extended discussions with the authors (\textbf{S-1}).
Polynomial approximations and exact positivity checks complete the
argument. The authors state that they wrote the manuscript and Lean
theorem statements, while Codex generated formal proofs under human
supervision. The reported formal verification includes the finite
computations, supporting \textbf{A-3}/\textbf{N-3}, and the author-written
exposition supports \textbf{T-2}. The case illustrates how AI-assisted
resolution of a key inequality can be integrated into a human-oriented
exposition and a formally verified proof. 

\paragraph{Courtade--Kumar and Hellinger results in OpenAI's collection.} OpenAI’s public mathematics collection \cite{openai_math_2026} also includes two related manuscripts, \emph{Sharp binary-information contraction on the discrete cube} and \emph{Hellinger contraction with arbitrary Boolean output bias}. These report a Courtade--Kumar result and a proof of the full Hellinger conjecture, respectively, and should be distinguished from the weak Hellinger result discussed above. The repository describes the manuscripts as outputs of an internal AI model.  Its accompanying Lean scope note documents formalization of the binary-information contraction result \cite{openai_math_119_2026}, including the Courtade--Kumar inequality, but not the proof for the full Hellinger inequality. 
We assess the two manuscripts separately. The documented Lean
formalization supports \textbf{A-3} for the binary-information
contraction result. For the full Hellinger argument, a complete
human analytical check or formal verification is not discussed
in the materials examined, and we leave \textbf{A}-level unassigned. The remaining
classifications require separate assessment of the disclosed
workflow, computational obligations, and preparation of the
exposition. For the named conjectures, we assign \textbf{F-0}.
The collection describes the manuscripts as model-produced outputs,
with no substantive human reconstruction for these
two cases, supporting \textbf{T-0}. The disclosed evaluation
procedure does not sufficiently specify the interaction or
orchestration mechanism to assign an \textbf{S}-level.

\section{Comparison and Observations}

\subsection{Comparison}

Table~\ref{tab:pipeline} summarizes the documented research
pipeline. The columns record problem formulation (\textbf{F}),
AI-mediated solution (\textbf{S}), analytical verification
(\textbf{A}), numerical certification (\textbf{N}), and translation
(\textbf{T}). A dash means that the public description does not
support a specific classification. The final column gives the qualitative reasoning/computation profile.
Reasoning-dominant and
computation-dominant indicate the predominant mode, while mixed
indicates substantial contributions from both, without implying
equal shares. It
does not measure the division of work or credit between humans and
AI.

\begin{table}[p]
\centering
\small
\setlength{\tabcolsep}{3pt}
\renewcommand{\arraystretch}{1.08}


\begin{tabular}{
@{}>{\raggedright\arraybackslash}p{0.46\textwidth}
cccccc@{}}
\hline
\hline
\textbf{Result}
& \textbf{F}
& \textbf{S}
& \textbf{A}
& \textbf{N}
& \textbf{T}
& \textbf{Remark}\\
\hline\hline

Gaussian monotonicity counterexample \cite{gu_sellke_2026}
& \textbf{F-0} & -- & -- & \textbf{N-2}
& \textbf{T-2}$^{*}$ & \makecell[c]{Reasoning-dominant}\\

Fisher-information counterexample \cite{zou_et_al_2026}
& \textbf{F-0} & -- & -- & \textbf{N-0}
& \textbf{T-2}$^{*}$ & \makecell[c]{Reasoning-dominant}\\

Multiple unicast: Reed--Solomon construction
\cite{braverman_he_2026}
& \textbf{F-0} & -- & \textbf{A-1} & \textbf{N-2}
& -- & \makecell[c]{Reasoning-dominant}\\

Multiple unicast: reversible circuits \cite{zhang_li_li_2026}
& \textbf{F-0} & -- & \textbf{A-3} & \textbf{N-0}
& -- & \makecell[c]{Reasoning-dominant}\\

DSBS channel conjectures \cite{pichler_2026}
& \textbf{F-0} & -- & \textbf{A-3}
& \makecell[c]{\textbf{N-0}\\\textbf{N-3}}
& -- & \makecell[c]{Reasoning-dominant}\\

Deletion-channel upper bound \cite{soysal_2026}
& \textbf{F-0} & -- & \textbf{A-3} & \textbf{N-3}
& -- & Mixed\\

Recursive zero-error constructions \cite{tandon_2026}
& \textbf{F-0} & -- & -- & \textbf{N-2}
& -- & Mixed\\

Odd-cycle capacity constructions \cite{itty_et_al_2026}
& \textbf{F-0} & -- & -- & \textbf{N-2}
& \textbf{T-2} & \makecell[c]{Computation-dominant}\\
\hline

Bernoulli affine-subspace entropy \cite{wu_bernoulli_2026}
& \textbf{F-0} & \textbf{S-1} & \textbf{A-1}
& \textbf{N-0} & \textbf{T-2} & \makecell[c]{Reasoning-dominant}\\

Marton inner-bound counterexample \cite{huang_liu_liu_2026}
& \textbf{F-0} & \textbf{S-1}
& \makecell[c]{\textbf{A-1}/\\\textbf{A-2}}
& \textbf{N-2} & \textbf{T-2}$^{*}$ & \makecell[c]{Reasoning-dominant}\\

Gaussian MAC: corner points \cite{tan_mac_corners_2026}
& \textbf{F-0} & \textbf{S-1}$^{*}$ & \textbf{A-1}
& \textbf{N-0} & \textbf{T-2}$^{*}$ & \makecell[c]{Reasoning-dominant}\\

Gaussian MAC: sum-rate face \cite{tan_mac_face_2026}
& \textbf{F-0} & \textbf{S-1}$^{*}$ & \textbf{A-1}
& \textbf{N-0} & \textbf{T-2}$^{*}$ & \makecell[c]{Reasoning-dominant}\\

Exact-repair regenerating codes \cite{tian_regenerating_2026}
& \textbf{F-0} & \textbf{S-1} & \textbf{A-1}
& \textbf{N-2} & \textbf{T-1} & Mixed\\

Rogers--Shephard entropy inequalities \cite{han_et_al_2026}
& \textbf{F-0} & \textbf{S-1}$^{*}$ & \textbf{A-1}
& \textbf{N-0} & \textbf{T-1}$^{*}$ & \makecell[c]{Reasoning-dominant}\\
\hline

Deletion-correcting codes \cite{weindel_heckel_2025}
& \textbf{F-0} & \textbf{S-2} & --
& \makecell[c]{\textbf{N-0}\\\textbf{N-2}}
& \textbf{T-2}$^{*}$ & \makecell[c]{Computation-dominant}\\

Deletion-channel capacity approximation \cite{papailiopoulos_2026}
& \textbf{F-0} & \textbf{S-2} & \textbf{A-0}$^{*}$
& \textbf{N-2} & \textbf{T-0} & Mixed\\

Fourier entropy of threshold functions \cite{shmalo_2026}
& \textbf{F-0} & \textbf{S-2} & --
& -- & -- & \makecell[c]{Reasoning-dominant}\\
\hline

Courtade--Kumar: entropy production and spectral arguments
\cite{ky_tran_2026}
& \textbf{F-0} & -- & -- & \textbf{N-2}
& \textbf{T-2}$^{*}$ & \makecell[c]{Reasoning-dominant}\\

Courtade--Kumar: differential equations and Bellman inequalities
\cite{chen_et_al_ck_2026}
& \textbf{F-0}
& \makecell[c]{\textbf{S-1}\\\textbf{S-2}}
& \textbf{A-3} & \textbf{N-3}
& \textbf{T-2}$^{*}$ & Mixed\\

Courtade--Kumar: entropy-flow proof
\cite{mahdavifar_beirami_2026}
& \textbf{F-0} & \textbf{S-1}$^{*}$
& \textbf{A-3} & \textbf{N-0}
& \textbf{T-0}$^{*}$ & \makecell[c]{Reasoning-dominant}\\

Multibit constructions and conjecture
\cite{mahdavifar_beirami_2026}
& \textbf{F-0} & \textbf{S-1}$^{*}$ & --
& \makecell[c]{\textbf{N-0}\\\textbf{N-2}\\\textbf{N-1}}
& \textbf{T-2}$^{*}$ & Mixed\\

Weak Hellinger inequality \cite{durcik_et_al_2026}
& \textbf{F-0} & \textbf{S-1} & \textbf{A-3}
& \textbf{N-3} & \textbf{T-2} & Mixed\\

OpenAI collection: binary-information contraction
\cite{openai_math_2026,openai_math_119_2026}
& \textbf{F-0} & -- & \textbf{A-3}
& -- & \textbf{T-0} & --\\

OpenAI collection: full Hellinger argument \cite{openai_math_2026}
& \textbf{F-0} & -- & --
& -- & \textbf{T-0} & --\\
\hline
\end{tabular}

\caption{Classification of the AI-assisted results discussed in
Section~4. F, S, A, N, and T refer to the taxonomy in Section~2.
An asterisk indicates an inferred assignment; a dash indicates
insufficient information for an assignment. The Remark column gives the qualitative reasoning/computation
profile. }
\label{tab:pipeline}

\end{table}

Several entries distinguish between claims within the same work.
For the multibit extension of Mahdavifar and Beirami
\cite{mahdavifar_beirami_2026}, \textbf{N-0} applies to the analytical
counterexamples for $k\geq 8$, \textbf{N-2} to the exact enumeration
within a specified family at $k=7$, and \textbf{N-1} to the broader
numerical evidence motivating the small-output conjecture.
For Weindel and Heckel \cite{weindel_heckel_2025}, the one-deletion
theorem is analytical (\textbf{N-0}), while rigorous finite checks
support other construction claims (\textbf{N-2}). Pichler's results
\cite{pichler_2026} similarly include proofs with no numerical
obligations (\textbf{N-0}) and a proof with formally integrated
computational certificates (\textbf{N-3}).

For the Marton counterexample \cite{huang_liu_liu_2026},
\textbf{A-1}/\textbf{A-2} reflects uncertainty about whether the
reported expert checking was independent of proof generation.
The inferred \textbf{A-0}$^{*}$ assessment for the deletion-channel
capacity proof \cite{papailiopoulos_2026} applies to the proof as
released and rests on the author's public account
\cite{papailiopoulos_blog_2026}. An unspecified A-level means that
the available information does not establish what analytical
verification took place.

The \textbf{T-2}$^{*}$ assignment for Chen et al.\
\cite{chen_et_al_ck_2026} applies to the accompanying expository
account. As elsewhere in the table, the asterisk marks an inference.
Formal verification establishes the formalized claim but does not
by itself establish how the exposition was prepared.

The reasoning/computation profiles describe the relative roles of
reasoning and computation in obtaining each result, including during
discovery. Thus, a result classified as \textbf{N-0} may involve
computation during discovery even when computation is unnecessary
in the final proof. Conversely, \textbf{N-3} describes how a
computational obligation was verified, regardless of how much
computation the work involved.

\subsection{Observations Across the Pipeline}

\paragraph{Problem formulation remains largely human-led.}
In the cases surveyed here, the consequential research question was
formulated or selected by human researchers. Indeed, every case in
Table~\ref{tab:pipeline} is classified as \textbf{F-0}. AI systems
have contributed to finding solutions and sometimes to sharpening
intermediate questions, but the available accounts do not establish
an AI-originated research problem in the sense of \textbf{F-2}.
This observation concerns the present sample, not a limit on what
AI systems may be able to formulate in the future. Whether an author is willing to disclose such information is also a concern in the case of \textbf{F-2}.

\paragraph{Human articulation is common, but its future form is open.}
Our classifications indicate substantial human involvement in articulation in many of the surveyed cases. Authors have converted
AI-assisted discoveries into proofs, explained constructions, and
distinguished established claims from numerical evidence and
conjectures. Such work can expose gaps and, more broadly, make a
result intelligible to other researchers. 

It is less clear whether comparable exposition will always require
authors to perform this translation themselves. An author might
instead direct an AI system, through a sequence of prompts, to
explain the underlying ideas and organize the final presentation.
We would classify a substantively guided and reviewed process of
this kind as \textbf{T-1}. The resulting text might be clear, but
its clarity alone would not show how well the authors understand
the argument or whether the exposition faithfully represents the
verified result. Those questions should be addressed separately
from the quality of the prose.

\paragraph{Solution and verification practices are more varied.}
The cases exhibit direct model assistance, extended mathematical
interaction, evolutionary search, and coding-agent workflows.
They also differ in how analytical arguments and finite computations
are checked. These choices reflect the structure of the problem,
the suitability of available tools, and the researchers' methods
and preferences. Consequently, an account of AI involvement is
most informative when it identifies the solution mechanism and
the verification performed for the resulting claims, rather than
assigning a single label to the paper. 

Several recent works supply  \textbf{A-3} level verifications; however, such a formalization requires a verification of the translation between the intended results and formalized results, particularly when the formalization is AI-generated. Our classification does not include a separate class for whether this verification is performed, which may be required for more complex proofs. 

\section{Conclusion}

The cases surveyed here show several ways in which AI can assist
information-theoretic research, rather than a single model of
AI-generated mathematics. At present, human researchers still
formulate or select the consequential questions in the documented
cases. The solution stage is more varied, ranging from model-assisted
counterexample discovery and mathematical dialogue to harnessed
search and agentic proof production. A single paper may also combine
human- and AI-developed components that deserve separate description.

Verification must follow the logical structure of each claim.
Certification of finite computations does not establish the
analytical reduction that makes them relevant, and corroboration
of a theorem by another proof does not check a particular
derivation. Publications should therefore state who verified the
indispensable analytical steps and what the numerical certificates
establish.

Most works in this sample also involve substantial human effort in
articulating their results. AI may increasingly perform parts of
that translation under authors' guidance, but a clear final text
alone does not establish human understanding or endorsement of its
claims. Scientific communication should preserve both an auditable
research record and an account that enables people to understand
the ideas, evidence, and limitations. Clear disclosure of AI's role
throughout the pipeline will help researchers and the broader
public assess what was discovered, how it was checked, and what
human experts can explain and stand behind.

Our assessments are based on public descriptions, which may be incomplete,
and some classifications necessarily involve interpretation. They should
therefore be revised as additional disclosures and verification evidence
become available. The taxonomy itself may also require refinement as AI
capabilities and research practices evolve. We offer this survey as a
starting point for continued discussion and invite additional cases,
corrections, and updates through the accompanying website. An accompanying webpage 
(\url{https://ct2641.github.io/AI_IT/}) collects the cases and classifications. Additional cases and corrections
can be submitted through GitHub Issues at
\url{https://github.com/ct2641/AI_IT/issues}.

\section*{Disclosure of AI Usage}

The authors established the scope and central perspective of this article.
AI systems were used interactively to assist with identifying and
summarizing relevant literature, proposing and refining classifications,
examining the consistency of the framework, and drafting and revising
portions of the manuscript. The authors directed this process and revised
the resulting text.

\section*{Acknowledgement}
The authors thank Amin Gohari, Chandra Nair, Hessam Mahdavifar, and Vincent Tan for their comments and suggestions, which helped improve the manuscript. 

\bibliographystyle{ieeetr}
\bibliography{ai_information_theory}

@article{gu_sellke_2026,
  author       = {Gu, Yuzhou and Sellke, Mark},
  title        = {A Counterexample to the Gaussian Completely Monotone Conjecture},
  year         = {2026},
  journal      = {arXiv preprint arXiv:2605.11656},
  eprint       = {2605.11656},
  archivePrefix= {arXiv},
  primaryClass = {math.PR},
  url          = {https://arxiv.org/abs/2605.11656}
}

@article{tan_mac_corners_2026,
  author        = {Tan, Vincent Y. F.},
  title         = {Second-Order Asymptotics for the {Gaussian} Multiple-Access
                   Channel at Corner Points},
  year          = {2026},
  journal      = {arXiv preprint arXiv:2608.18231},
  eprint        = {2608.18231},
  archivePrefix = {arXiv},
  primaryClass  = {cs.IT},
  url           = {https://arxiv.org/abs/2608.18231}
}

@article{tan_mac_face_2026,
  author        = {Tan, Vincent Y. F.},
  title         = {Second-Order Asymptotics for the {Gaussian} {MAC} on the
                   Relative Interior of the Sum-Rate Face},
  year          = {2026},
  journal      = {arXiv preprint arXiv:2609.18013},
  eprint        = {2609.18013},
  archivePrefix = {arXiv},
  primaryClass  = {cs.IT},
  url           = {https://arxiv.org/abs/2609.18013}
}

@article{zou_et_al_2026,
  author       = {Zou, Jiayang and Fan, Luyao and Gao, Jiayang and Wang, Jia},
  title        = {A Hexagonal Counterexample to Log-Convexity of {Fisher} Information Along the Heat Flow},
  year         = {2026},
  journal      = {arXiv preprint arXiv:2605.18081},
  eprint       = {2605.18081},
  archivePrefix= {arXiv},
  primaryClass = {math.PR},
  url          = {https://arxiv.org/abs/2605.18081}
}

@article{cheng2015higher,
  title={Higher order derivatives in {Costa’s} entropy power inequality},
  author={Cheng, Fan and Geng, Yanlin},
  journal={IEEE Transactions on Information Theory},
  volume={61},
  number={11},
  pages={5892--5905},
  year={2015},
  publisher={IEEE}
}

@article{marton1979coding,
  title={A coding theorem for the discrete memoryless broadcast channel},
  author={Marton, Katalin},
  journal={IEEE Transactions on Information Theory},
  volume={25},
  number={3},
  pages={306--311},
  year={1979},
  publisher={IEEE}
}

@article{tao2026mathematics,
  title={Mathematics in the age of {AI}},
  author={Tao, Terence},
  journal={arXiv preprint arXiv:2608.16753},
  year={2026}
}

@article{wu_bernoulli_2026,
  author       = {Wu, Yihong},
  title        = {Entropy of {Bernoulli} Measures Conditioned on Affine Subspaces and a Problem of {Ancheta--Massey}},
  year         = {2026},
  journal      = {arXiv preprint arXiv:2608.22837},
  eprint       = {2608.22837},
  archivePrefix= {arXiv},
  primaryClass = {cs.IT},
  url          = {https://arxiv.org/abs/2608.22837}
}

@article{huang_liu_liu_2026,
  author       = {Huang, Mian and Liu, Yanxiao and Liu, Yi},
  title        = {Sub-optimality of {Marton's} Inner Bound for the Two-Receiver Broadcast Channel},
  year         = {2026},
  journal      = {arXiv preprint arXiv:2608.19869},
  eprint       = {2608.19869},
  archivePrefix= {arXiv},
  primaryClass = {cs.IT},
  url          = {https://arxiv.org/abs/2608.19869}
}

@article{weindel_heckel_2025,
  author       = {Weindel, Franziska and Heckel, Reinhard},
  title        = {{LLM}-Guided Search for Deletion-Correcting Codes},
  year         = {2025},
  note         = {Revised 2026},
  journal      = {arXiv preprint arXiv:2504.00613},
  eprint       = {2504.00613},
  archivePrefix= {arXiv},
  primaryClass = {cs.IT},
  url          = {https://arxiv.org/abs/2504.00613}
}

@article{papailiopoulos_2026,
  author       = {Papailiopoulos, Dimitris},
  title        = {Binary Deletion Channel Capacity to Within One Hundredth of a Bit},
  year         = {2026},
  journal      = {arXiv preprint arXiv:2609.19412},
  eprint       = {2609.19412},
  archivePrefix= {arXiv},
  primaryClass = {cs.IT},
  url          = {https://arxiv.org/abs/2609.19412}
}

@misc{papailiopoulos_repo_2026,
  author       = {Papailiopoulos, Dimitris},
  title        = {Binary Deletion Channel Capacity: Code and Certificates},
  year         = {2026},
  howpublished = {GitHub repository},
  url          = {https://github.com/anadim/binary-deletion-channel-capacity},
  note         = {Accessed 24 September 2026}
}

@article{mahdavifar_beirami_2026,
  author       = {Mahdavifar, Hessam and Beirami, Ahmad},
  title        = {The Most Informative Bit and Beyond: A Proof of the {Courtade--Kumar} Conjecture and Multibit Extensions},
  year         = {2026},
  journal      = {arXiv preprint arXiv:2609.26444},
  eprint       = {2609.26444},
  archivePrefix= {arXiv},
  primaryClass = {cs.IT},
  url          = {https://arxiv.org/abs/2609.26444}
}

@article{chen_et_al_ck_2026,
  author       = {Chen, Zijie and Gohari, Amin and Javanmard, Adel and Lin, Honghao and Mirrokni, Vahab and Nair, Chandra and Woodruff, David P.},
  title        = {A Proof of the Most Informative {Boolean} Function Conjecture},
  year         = {2026},
  journal      = {arXiv preprint arXiv:2609.24931},
  eprint       = {2609.24931},
  archivePrefix= {arXiv},
  primaryClass = {cs.IT},
  url          = {https://arxiv.org/abs/2609.24931}
}

@article{ky_tran_2026,
  author       = {Ky, Vu Khac and Tran, Tuan},
  title        = {Dictators Are Most Informative},
  year         = {2026},
  journal      = {arXiv preprint arXiv:2609.24184},
  eprint       = {2609.24184},
  archivePrefix= {arXiv},
  primaryClass = {math.PR},
  url          = {https://arxiv.org/abs/2609.24184}
}

@article{braverman_he_2026,
  author = {Mark Braverman and Zhongtian He},
  title = {The Multiple Unicast Conjecture Is False},
  journal = {arXiv preprint arXiv:2610.10108},
  year = {2026},
  url = {https://arxiv.org/abs/2610.10108}
}

@article{tian_regenerating_2026,
  author = {Chao Tian},
  title = {Exact Storage-Repair-Bandwidth Tradeoffs for $(6,5,5)$ and $(7,6,6)$ Exact-Repair Regenerating Codes},
  journal = {arXiv preprint arXiv:2610.28199},
  year = {2026},
  url = {https://arxiv.org/abs/2610.28199}
}

@article{han_et_al_2026,
  author = {Bang-Xian Han and Guoqing Zhan and Chiyu Zhou and Zhuo-Nan Zhu},
  title = {Sharp {Rogers--Shephard} Inequalities for {R\'enyi} Entropy},
  journal = {arXiv preprint arXiv:2610.15716},
  year = {2026},
  url = {https://arxiv.org/abs/2610.15716}
}

@article{durcik_et_al_2026,
  author = {Polona Durcik and Marco Fraccaroli and Joris Roos},
  title = {A Weak {Hellinger} Inequality for Noisy {Boolean} Channels},
  journal = {arXiv preprint arXiv:2609.28534},
  year = {2026},
  url = {https://arxiv.org/abs/2609.28534}
}

@article{pichler_2026,
  author = {Georg Pichler},
  title = {Three Conjectures on Binary Channels for the Doubly Symmetric Binary Source},
  journal = {arXiv preprint arXiv:2609.27991},
  year = {2026},
  url = {https://arxiv.org/abs/2609.27991}
}

@article{soysal_2026,
  author = {\"Ozg\"ur Soysal},
  title = {A New Upper Bound on the Binary Deletion Channel Capacity},
  journal = {arXiv preprint arXiv:2609.13351},
  year = {2026},
  url = {https://arxiv.org/abs/2609.13351}
}

@article{shmalo_2026,
  author = {Yitzchak Shmalo},
  title = {{Fourier} Entropy of Threshold Functions and Their Compositions},
  journal = {arXiv preprint arXiv:2610.16069},
  year = {2026},
  url = {https://arxiv.org/abs/2610.16069}
}

@article{itty_et_al_2026,
  author = {Nathaniel Itty and Christopher D. Rosin and Chase Carstensen and Daniel Reichman},
  title = {Improved Lower Bounds for the {Shannon} Capacity of Odd Cycles},
  journal = {arXiv preprint arXiv:2607.21517},
  year = {2026},
  url = {https://arxiv.org/abs/2607.21517}
}

@article{tandon_2026,
  author = {Ravi Tandon},
  title = {Strengthening Recursive Constructions for Zero-Error {Shannon} Capacity},
  journal = {arXiv preprint arXiv:2608.30273},
  year = {2026},
  url = {https://arxiv.org/abs/2608.30273}
}

@inproceedings{li_li_2004,
  author    = {Li, Zongpeng and Li, Baochun},
  title     = {Network Coding: The Case of Multiple Unicast Sessions},
  booktitle = {Proceedings of the 42nd Annual Allerton Conference
               on Communication, Control, and Computing},
  year      = {2004},
  address   = {Monticello, IL, USA},
  url       = {https://iqua.ece.utoronto.ca/papers/allerton04.pdf}
}

@article{zhang_li_li_2026,
  author  = {Zhang, Xindan and Li, Baochun and Li, Zongpeng},
  title   = {Network Coding Can Beat Routing in Undirected
             Multiple-Unicast Networks},
  journal = {arXiv preprint arXiv:2610.09367},
}

@misc{openai_math_2026,
  author = {{OpenAI}},
  title  = {{OpenAI} Mathematics Collection},
  year   = {2026},
  howpublished = {GitHub repository},
  note   = {\url{https://github.com/openai/math}.
            Accessed November 2, 2026}
}

@misc{openai_math_119_2026,
  author = {{OpenAI}},
  title  = {The {Courtade--Kumar} and {Hellinger} Conjectures:
            Scope of the {Lean} Formalization},
  year   = {2026},
  howpublished = {Documentation accompanying the {OpenAI} mathematics collection},
  note   = {\url{https://github.com/openai/math/blob/main/lean/docs/119.md}.
            Accessed November 2, 2026}
}

@misc{papailiopoulos_blog_2026,
  author = {Papailiopoulos, Dimitris},
  title = {One Hundredth of a Bit},
  year = {2026},
  month = sep,
  howpublished = {Substack, September 15, 2026},
  note = {\url{https://dimitrisp.substack.com/p/one-hundredth-of-a-bit}.
          Accessed November 4, 2026}
}

\end{document}